\documentclass[twocolumn,showpacs,preprintnumbers,amsmath,amssymb,superscriptaddress]{revtex4}

\usepackage{graphicx}
\usepackage{dcolumn}
\usepackage{bm}
\usepackage{bigints}
\usepackage{xcolor}  

\begin{document}
\title{Non-Markovian dynamics of a single-mode cavity strongly coupled to an inhomogeneously broadened spin ensemble} 

\author{Dmitry O. Krimer}
\email[]{dmitry.krimer@gmail.com}
\affiliation{Institute for Theoretical Physics, Vienna University of Technology, Wiedner Hauptstrasse 8-10/136, 1040 Vienna, Austria}
\author{Stefan Putz}
\affiliation{Vienna Center for Quantum Science and Technology, Atominstitut, Vienna University of Technology, Stadionallee 2, 1020 Vienna, Austria}
\author{Johannes Majer}
\affiliation{Vienna Center for Quantum Science and Technology, Atominstitut, Vienna University of Technology, Stadionallee 2, 1020 Vienna, Austria}
\author{Stefan Rotter}
\affiliation{Institute for Theoretical Physics, Vienna University of Technology, Wiedner Hauptstrasse 8-10/136, 1040 Vienna, Austria}

\begin{abstract}
We study the dynamics of a spin ensemble strongly coupled to a single-mode resonator driven by external pulses. When the mean frequency of the spin ensemble is in resonance with the cavity mode, damped Rabi oscillations are found between the spin ensemble and the cavity mode which we describe very accurately, including the dephasing effect of the inhomogeneous spin broadening. We demonstrate that a precise knowledge of this broadening is crucial both for a qualitative and a quantitative understanding of the temporal spin-cavity dynamics. On this basis we show that coherent oscillations between the spin ensemble and the cavity can be enhanced by a few orders of magnitude, when driving the system with pulses that match special resonance conditions. Our theoretical approach is tested successfully with an experiment based on an ensemble of negatively charged nitrogen-vacancy (NV) centers in diamond strongly coupled to a superconducting coplanar single-mode waveguide resonator.
\end{abstract}
  
 
\maketitle

\section{Introduction}

\begin{figure}[h!]
\includegraphics[angle=0,width=.85\columnwidth]{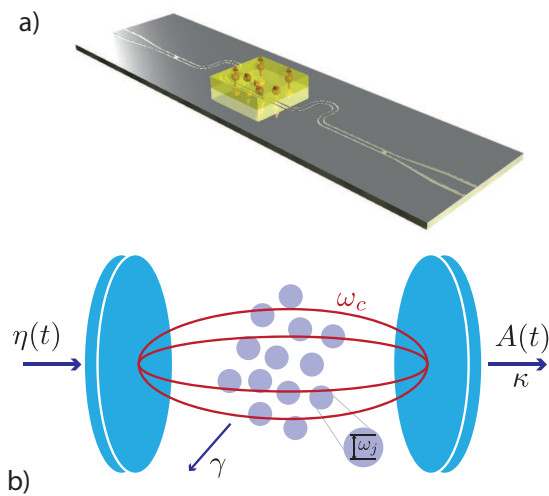}
\caption{(Color online) Sketch of the hybrid quantum system studied in this paper: a) a spin ensemble (yellow) coupled to a transmission-line resonator (gray) confining the electromagnetic field inside a small volume. b) Scheme of the spin ensemble-cavity coupled system. An incoming signal $\eta(t)$ passes through the cavity characterized by a frequency $\omega_c$ which is coupled to a spin ensemble with each individual spin of frequency $\omega_j$. The transmitted signal is proportional to the cavity amplitude, $A(t)$. $\kappa$ and $\gamma$ stands for the  cavity and spin losses, respectively.}
\label{Fig1_ensemble_cavity}
\end{figure}

Over the past decade various setups in cavity quantum electrodynamics (QED) have been studied in terms of their potential for future technologies involving the storage and processing of quantum information. Among different hybrid quantum systems \cite{You2005}, the ones based on spin, atomic or even molecular ensembles coupled to superconducting microwave cavities have recently attracted much attention \cite{Amsuess2011,Sandner2012,Kubo2010,Kubo2011,Kubo2012,Probst2013,Rabl2006,Verdu2009,Xiang2013}, see Fig.~\ref{Fig1_ensemble_cavity}. In such systems the spin or atomic ensemble plays the role of a quantum memory, to which the quantum information is coherently stored and retrieved from at some later time. The cavity, in turn, serves as a quantum bus for the in- and output of information as well as for the coupling between several constituents of such hybrid quantum systems (see e.g. \cite{Kubo2011}). One of the necessary conditions for the coherent transfer of quantum information between an ensemble and a cavity is the strong coupling between them. Fortunately, various spin ensembles, as for instance, negatively charged nitrogen-vacancy (NV) defects in diamond \cite{Amsuess2011,Sandner2012,Kubo2010,Kubo2011,Kubo2012}, rare-earth spin ensembles \cite{Probst2013}, clouds of ultracold atoms \cite{Atac2009,Verdu2009} or magnons in yttrium iron garnet with or without doping \cite{Huebl2013,Tabuchiarxiv}, may satisfy this requirement when being collectively coupled to \cite{Dicke54}. We also note that in recent proposals the direct coupling of a qubit to such spin-ensembles has been suggested without any cavity being involved \cite{Zhu2011,Saito2013}. 

Here we study the dynamics of a superconducting cavity strongly coupled to an ensemble of negatively charged NV centers. Each individual NV center can possess a sufficiently long coherence time \cite{Bar-Gill2013} needed for the coherent transfer of quantum information. However, since the local magnetic dipole-dipole couplings of NV centers constituting the ensemble to the bath of magnetic impurities (such as nitrogen atoms not converted into NV centers) slightly differ from each other, the NV electron spin resonance line of a large ensemble is inhomogeneously broadened \cite{Stanwix2010}. This line broadening acts as the main source of decoherence, and constitutes a significant drawback of this solid-state spin ensemble leading to a drastic decrease of its coherence time. Several approaches including echo-type refocusing techniques \cite{Grezes2014,Wu2010} have meanwhile been suggested to overcome this limitation. Recent stationary transmission studies demonstrate that the decoherence can be strongly suppressed altogether \cite{Kurucz2011,Diniz2011} when the spin density has a spectral distribution with tails that decay sufficiently fast \cite{Kurucz2011,Diniz2011,Sandner2012}. In this paper we report on a detailed time-dependent study for exactly such a case and demonstrate how the corresponding dynamics can be efficiently captured using a Volterra integral equation for the cavity amplitude \cite{Nature2014}. The excellent correspondence between our theoretical model and a corresponding experiment allows us to closely look into the fascinating features following from a pulsed driving of this hybrid quantum system in the strong-coupling regime.

Our paper is organized as follows. In section II we present the theoretical framework of our problem and summarize the most important assumptions made. We sketch the general form of the equations obtained, describing the two methods for solving the Volterra equation in Appendices \ref{App_time_integr_Volt_Eq}, \ref{App_Lapl_Transf}.
Furthermore, we discuss the specific experimental realization of our theory. In section III, we consider the dynamics under the action of a long rectangular microwave pulse which allows us to obtain the precise form for the spin density and its parameters by detailed comparison with the experimental results. We also present analytical results for a Lorentzian spin density distribution and demonstrate which features are captured by this approximation and which are not. Section IV will then address the question how the decoherence in our system caused by inhomogeneous broadening changes as a function of the coupling strength. We show that a non-Lorentzian functional profile of the spin distribution leads to a strong suppression of decoherence for large values of the coupling strength -- an effect known as ``cavity protection'' \cite{Kurucz2011,Diniz2011}. Finally, in section V, we propose a scheme which allows us to induce giant coherent oscillations between the cavity and our spin ensemble as well as to transfer energy into the spin ensemble very efficiently.


%
\section{Theoretical model}
\label{Sec_Theory}

We study the temporal dynamics of a system consisting of a large spin ensemble coupled with a single-mode cavity via magnetic or electric dipole interaction. We assume that the distance between spins is large enough such that the dipole-dipole interactions between spins can be neglected. Our starting point is the Tavis-Cummings Hamiltonian ($\hbar=1$) \cite{Tavis68}
\begin{eqnarray}
&&{\cal H}=\omega_ca^{\dagger}a+\frac{1}{2}\sum_j^N\omega_j\sigma_j^z+\text{i}\sum_j^N\left[g_j\sigma_j^-a^{\dagger}-g_j^*\sigma_j^+a\right]-
\nonumber\\
&&\text{i}\left[\eta(t) a^{\dagger}\text{e}^{-\text{i}\omega_p t}-\eta(t)^* a\text{e}^{\text{i}\omega_p t}\right]\,,
\label{Hamilt_fun}
\end{eqnarray}
where $a^{\dag}$ and $a$ are standard creation and annihilation operators of the single cavity mode with frequency $\omega_c$ and $\sigma_j^+,\,\sigma_j^-,\,\sigma_j^z$ are the Pauli operators associated with each individual spin of frequency $\omega_j$. An incoming signal is characterized by the carrier frequency $\omega_p$ and by the amplitude $\eta(t)$ whose time variation is much slower as compared to $1/\omega_p$. The interaction part of ${\cal H}$ is written in the dipole and rotating-wave approximation (terms $\propto a\sigma_j^-,\,a^\dag \sigma_j^+$ are neglected), where $g_j$ stands for the coupling strength of the $j$-th spin.

Despite the fact that each individual spin is coupled weakly to the cavity, one can nevertheless reach the strong coupling regime due to the large number of spins which are collectively coupled to the cavity mode (see e.g. \cite{Amsuess2011,Verdu2009,Kubo2011} for NV spin ensembles). The effect of collective coupling is particularly  evident when reducing the interaction term to a collective term $\Omega(S^-a^{\dagger}-S^+a)$ \cite{Emary2003}, where the collective spin operators are given by $S^\pm=N^{-1/2}\cdot \sum_j^N\sigma_j^\pm$. The prefactor $\Omega^2=\sum_j^Ng_j^2$ stands for an effective coupling strength, which scales up a single coupling strength, $g_j$, by a factor of $\sqrt{N}$, so that $\Omega$ can be sufficiently enhanced for the realization of the strong coupling regime. In this formulation the effective spin-waves that are excited by the cavity mode can be identified as superradiant collective Dicke states which are effectively damped by the coupling to subradiant states in the ensemble \cite{Dicke54,Kurucz2011,Diniz2011}. Note that the rotating-wave approximation mentioned above is applicable only if 
$\Omega \ll \omega_c$.

Next, we derive the Heisenberg operator equations, for the cavity and spin operators, $\dot a=i [{\cal H},a]-\kappa a$, $\dot \sigma_k^-=i [{\cal H},\sigma_k^-]-\gamma \sigma_k^-$, respectively. Here $\kappa$ and $\gamma$ stand for the total dissipative cavity and spin losses.  Strictly speaking, the noise operators should also be added to the r.h.s. of these equations in order to preserve the commutation relations. However, their expectation values vanish 
as was shown already in earlier works \cite{Kurucz2011,Diniz2011} on the example of an NV ensemble and therefore these terms are not included here explicitly. These Heisenberg equations describe the dynamics to a very high accuracy, provided that the energy of photons of the external bath is substantially smaller than that of cavity photons, $kT\ll\hbar \omega_c$.  We then write a set of equations for the expectation values, $\langle a(t)\rangle$ and $\langle\sigma_k^-(t)\rangle$ in the frame rotating with the probe frequency $\omega_p$. In what follows the amplitude of the pumping signal $\eta(t)$ is taken to be rather small and therefore the number of the excited spins is always small compared to the ensemble size. This allows us to simplify these equations further by setting $\langle \sigma_k^z \rangle \approx -1$ (Holstein-Primakoff-approximation \cite{Primakoff1939}). With all these simplifications the equations for the cavity and spin amplitudes become
\begin{subequations}
\begin{eqnarray}
\label{Eq_a_Volt}
&&\!\!\!\!\!\!\!\!\!\!\!\!\!\!\!\! \dot{A}(t)= -\left[\kappa+i(\omega_c-\omega_p)\right]A(t) + \sum_k
g_k  B_k(t)-\eta(t), \\
\label{Eq_bk_Volt}
&&\!\!\!\!\!\!\!\!\!\!\!\!\!\!\!\!\dot{B}_k(t) = -\left[\gamma+i(\omega_k-\omega_p) \right]B_k(t) - g_k A(t),
\end{eqnarray}
\end{subequations}
where $A(t)\equiv \langle a(t)\rangle$ and $B_k(t)\equiv\langle\sigma_k^-(t)\rangle$. 

\subsection*{Experimental realization}
In the following, we will compare our theoretical model with one specific experimental realization, namely a $\lambda/2$ superconducting microwave coplanar waveguide resonator magnetically coupled with a spin ensemble of negatively charged NV centers in diamond. The corresponding  experiment is carried out in a standard dilution refrigerator with a synthetic diamond placed on top of a resonator cooled to millikelvin temperatures ($\sim25$\,mK) (see \cite{Nature2014} for more details). The concentration of NV centers in diamond is sufficiently low and the distance between spins is still large enough, so that the dipole-dipole interactions between spins is negligibly small justifying the assumption of our model. By applying an external magnetic field, two degenerate sub-ensembles, which can effectively be considered as a single sub-ensemble,  are brought into resonance with the cavity, whereas the other sub-ensembles make a slight dispersive contribution only and their influence is neglected here (see e.g. \cite{Nature2014,Amsuess2011,Sandner2012} for more details). The individual spins are distributed around the mean frequency $\omega_s=2\pi\cdot 2.6915$\,GHz, with the width $\Delta \ll \omega_s$, which is of the order of $10$\,MHz. The coupling strength of each individual spin with a cavity mode is typically of the order of $g_j/2\pi\sim10$\,Hz \cite{Verdu2009}. However, the effective coupling $\Omega$ is enhanced by a factor of $\sqrt{N}$ with the ensemble size $N\sim 10^{12}$, so that $\Omega$ can reach values as large as $10$\,MHz which is sufficient to reach the strong coupling regime. Note that the energy of thermal photons is substantially smaller than that of microwave photons, $kT\ll\hbar \omega_c$, resulting in an occupation probability of the ensemble in the ground state which is larger than $0.99$. In what follows, the cavity frequency was taken to be always equal to the spin mean frequency, $\omega_c=\omega_s=2\pi\cdot 2.6915$\,GHz. Therefore the inequality $\Omega \ll \omega_c$ always holds, such that the rotating-wave approximation is very well fulfilled. Note also that the spin dissipation is much smaller than the cavity dissipation, $\gamma \ll \kappa$, so that the former does not contribute to the dynamics realized in the experiment. We thus omitted $\gamma$ everywhere, except when necessary for the calculation of some integrals which would otherwise be singular.

\subsection*{Setting up the Volterra integral equation}

Owing to the large number of spins within the ensemble ($N\sim 10^{12}$), there are a lot of spins in each frequency subinterval around $\omega_s$ which make a non-negligible contribution to the dynamics. We can thus introduce a continuous spectral density as $\rho(\omega)=\sum_k g_k^2 \delta(\omega-\omega_k)/\Omega^2$, where $\Omega^2=\sum_j^Ng_j^2$ is the collective coupling strength of the spin ensemble to the cavity, satisfying the normalization condition $\int d\omega\rho(\omega)=1$. As we shall see below, one should take special care when choosing the functional profile of the spectral distribution for the spin density, $\rho(\omega)$, which describes its inhomogeneous broadening and which plays a crucial role for the dynamics. 

To go to the continuous limit (in frequency) we carry out the following formal replacement from the discrete function $F(\omega_k)$ to the continuous one, $F(\omega)$: $\sum_k F(\omega_k) \rightarrow \Omega^2 \int d\omega \rho(\omega) F(\omega)$.  By integrating Eq.~(\ref{Eq_bk_Volt}) in time, each individual spin amplitude, $B_k(t)$, can be expressed in terms of the cavity amplitude, $A(t)$, as
\begin{eqnarray}
\label{Eq_Bk}
&&B_k(t)\!=\!B_k(0)e^{-i (\omega_k-\omega_p-i\gamma)t}\!-
\\\nonumber\!
&&g_k \int\limits_{0}^t\! d\tau e^{-i (\omega_k-\omega_p-i\gamma)(t-\tau)}\!\cdot\!A(\tau),
\end{eqnarray}
where $B_k(0)$ is the initial spin amplitude. Substituting Eq.~(\ref{Eq_Bk}) into Eq.~(\ref{Eq_a_Volt}) we arrive at the Volterra equation for the cavity amplitude, $A(t)$
\begin{eqnarray}
\nonumber
&&\!\!\!\!\!\!\!\!\!\dot A(t)\!=\!-i(\omega_c-\omega_p-i\kappa) A(t)\!+\!\sum_k g_k B_k(0) e^{-i(\omega_k-\omega_p-i\gamma)t}
\\
&&\!\!\!\!\!\!\!\!\!-\Omega^2 \int_0^{\infty} d\omega \rho(\omega) \int\limits_{0}^t d\tau
e^{-i(\omega-\omega_p-i\gamma)(t-\tau)}A(\tau)-\eta(t).
\label{Eq_a_with_Bk0}
\end{eqnarray}
After integrating Eq.~(\ref{Eq_a_with_Bk0}) in time, performing lengthy but straightforward algebraic calculations and assuming that the cavity is initially empty, $A(0)=0$, and all spins are initially in the ground state, $B_k(0)=0$, we end up with the following Volterra equation for the cavity amplitude
\begin{eqnarray}
\label{Volt_eq}
A(t)=\int\limits_0^t d\tau {\cal K}(t-\tau) A(\tau)+{\cal F}(t),
\end{eqnarray}
which contains the kernel function ${\cal K}(t-\tau)$,
\begin{eqnarray}
\label{Volt_eq_K}
&&\!\!\!\!\!\!\!\!{\cal K}(t-\tau)=\Omega^2\cdot
\\\nonumber
&&\!\!\!\!\!\!\!\! \bigintsss_0^{\infty}\!\!\! d\omega\,
\dfrac{\rho(\omega) \left[e^{-i (\omega-\omega_c+i \kappa)(t-\tau)}-1\right]
}{i (\omega-\omega_c+i \kappa)}\!\cdot\! e^{-i(\omega_c-\omega_p-i\kappa)(t-\tau)},
\end{eqnarray}
and the function ${\cal F}(t)$,
\begin{eqnarray}
\label{Volt_eq_F}
{\cal F}(t)=-\int\limits_0^t d\tau\, \eta(\tau)\cdot e^{-i(\omega_c-\omega_p-i\kappa)(t-\tau)},
\end{eqnarray}
where the amplitude, $\eta(t)$, represents an arbitrarily shaped incoming pulse or a sequence of pulses. Note, that the kernel function ${\cal K}(t-\tau)$ accounts for memory effects and leads in general to a non-Markovian feedback of the NV ensemble on the cavity. In Appendices \ref{App_time_integr_Volt_Eq}, \ref{App_Lapl_Transf} we give a detailed description of the two methods which allow us to solve the Volterra equation in a very efficient way. 

Having calculated the cavity amplitude, $A(t)$, we can find the expectation values of the collective spin operator, $J_x+i J_y=\sum_k g_k B_k(t)/[2(\sum_i g_i^2)^{1/2}]$, which in the continuous limit and for the initial conditions $A(0)=0$ and $B_k(0)=0$ introduced above read as follows
\begin{eqnarray}
\label{Eq_a_with_A0}
J_x+i J_y=-\dfrac{\Omega}{2} \int_0^{\infty} \! d\omega \rho(\omega) \int\limits_{0}^t  \! d\tau
e^{-i(\omega-\omega_c)(t-\tau)}A(\tau).
\end{eqnarray}
The $z$-component of the expectation value of the collective spin operator, $J_z=\sum_k \langle \sigma_k^z \rangle/(2\sqrt{N})$, remains $J_z \approx -\sqrt{N}$, in accordance with the approximations discussed above.

Note that Eqs.~(\ref{Eq_a_Volt},\ref{Eq_bk_Volt}), as well as the resulting Volterra equation (\ref{Eq_a_with_Bk0}) are linear equations with respect to the cavity and spin amplitudes, $A(t)$ and  $B_k(t)$, respectively. We can thus always rescale our solution by multiplying the amplitude of the driving signal, $\eta(t)$, by an arbitrary scaling factor. In the following we take the amplitude of the incoming signal equal to the cavity decay rate, $\eta=\kappa$. Note that such a choice corresponds to the situation when the incoming signal, being in a coherent state, gives rise to a single photon in the empty cavity on average. The experimental curves will be appropriately rescaled with a constant prefactor such as to match the corresponding theoretical curves.

\section{Dynamics under the action of a long pulse}
\label{Sec_dyn_long_pulse}
%
%
\begin{figure}
\includegraphics[angle=0,width=1.\columnwidth]{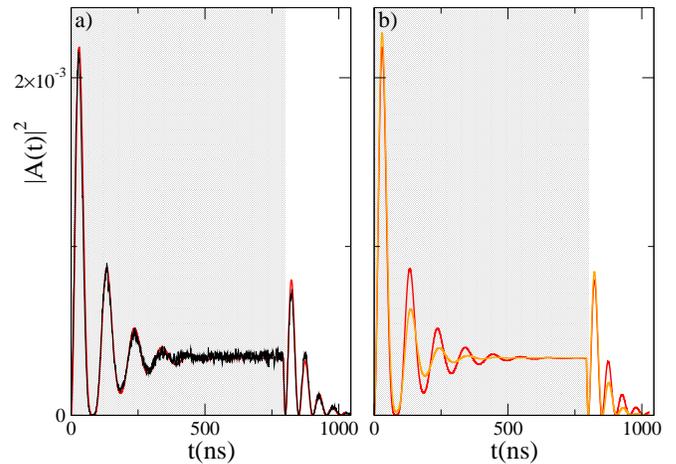}
\caption{(color online). Cavity probability amplitude $|A(t)|^2$ versus time $t$ under the action of an incident long rectangular pulse of duration $800$\,ns with the carrier frequency matching the resonance condition, $\omega_p=\omega_c=2\pi\cdot 2.6915$\,GHz, where $\omega_c$ stands for the cavity resonance frequency. Gray (white) area indicates a time interval during which the pumping signal is on (off). a) (taken from \cite{Nature2014}) Red (gray) curve: Numerical results for the cavity transmission at a coupling strength $\Omega/2\pi=8.56$~MHz. In the calculations the spectral density is modelled by a $q$-Gaussian distribution. The frequency of Rabi oscillations, $\Omega_R=2\pi\cdot 19.2$\,MHz. Black curve: experimental results for the cavity transmission. b) Red (gray) curve the same as in a). Orange (light gray) curve: results of numerical calculations assuming a Lorentzian distribution of the spin density.}
\label{Figure_long_pulse_with_Lorentz}
\end{figure}

In order to choose an appropriate form for the spectral density, $\rho(\omega)$, we compare our numerical results with the experiment performed within the strong-coupling regime. Specifically, we apply a rectangular microwave pulse [$\eta(t)=\eta$ for $0\le t \le \tau_d$ and $\eta(t)=0$ otherwise, see Eq.~(\ref{Volt_eq_F})], with the resonance carrier frequency ($\omega_p=\omega_c=\omega_s$). This pulse has a duration $\tau_d$ substantially longer than the resulting period of damped Rabi oscillations and the inverse of the total decay rate, so that the system sets into a steady state before the signal is turned off [see Fig.~\ref{Figure_long_pulse_with_Lorentz}a)]. Note that the total decay rate describes the overall decoherence in our system which consists of two contributions: The first one is due to dissipative cavity losses $\kappa$, while the second one originates from the inhomogeneous broadening of the spin ensemble which leads to the dephasing of spins during the time evolution. As we shall see below, this dephasing mechanism gives the dominant contribution to the decoherence (the spin dissipation $\gamma$ is negligible in our case).

In accordance with our previous study \cite{Sandner2012,Nature2014}, we obtain a very good agreement between theory and experiment, when taking a $q$-Gaussian \cite{Tsallis09} as the distribution function for the spectral density defined as
\begin{eqnarray}
\label{rho_w_Eq}
\rho(\omega)=C\cdot\left[1-(1-q)\dfrac{(\omega-\omega_s)^2}{\Delta^2}\right]^{
\dfrac{1}{1-q}}.
\end{eqnarray}
 Here $q$ is the dimensionless shape parameter, $1<q<3$, $\gamma_q=2\Delta\sqrt{\dfrac{2^q-2}{2q-2}}$ is the full-width at half maximum (FWHM) and $C$ is the normalization constant. Note, that for $q\rightarrow 1$ and  $q=2$ we recover a Gaussian  and Lorentzian distribution, respectively. From the comparison with the experiment, we extracted the following parameters used in our calculations: $q=1.39$, $\gamma_q/2\pi=9.4$\,MHz, and $\kappa/2\pi=0.8$\,MHz (FWHM of the cavity decay). We have also tested other lineshapes for describing the spectral spin density such as the stable alpha-distribution, but found them to be less suitable for describing the experimentally observed data.

An interesting and, at first sight, surprising fact is that the first Rabi peak of the cavity amplitude after switching off the microwave signal is approximately twice as large as the steady state amplitude, as seen in Fig.~\ref{Figure_long_pulse_with_Lorentz}a). This overshoot effect takes place after the incoming signal is turned off, because the energy stored in the spin ensemble is released back to the cavity and interferes constructively with the energy stored there (see  Appendix \ref{App_Decay} for more details). It will be shown in the next section that this overshoot appears only if the coupling strength is larger than a certain critical value. In addition to this condition, the overshoot effect also requires a finite amount of energy being stored in the spin ensemble, but does not show up if it is in the ground state and the field inside the cavity is described by a Fock state, as for instance when it is fed with a single photon, see Appendix \ref{App_Lapl_Transf}.

\subsection{Dynamics for a Lorentzian spin density distribution}
\label{Subsec_Dyn_Lorentz_distr}

To illustrate the importance of the spectral spin distribution, we have also tried to achieve an agreement with the experiment when assuming a Lorentzian instead of a $q$-Gaussian distribution for the spectral density, 
\begin{eqnarray}
\label{rho_Lorentz}
\rho(\omega)=\dfrac{\Delta}{\pi[(\omega-\omega_s)^2+\Delta^2]}.
\end{eqnarray}
For this purpose, we adapt the parameters such that the period of the resulting Rabi oscillations and the cavity amplitude at the steady-state agree with the measurements, see Fig.~\ref{Figure_long_pulse_with_Lorentz}b). As seen there, the Lorentzian predicts a sufficiently larger decay rate as compared to that observed in the experiment [compare the values of the Rabi peaks during damped Rabi oscillations for the $q$-Gaussian and for the Lorentzian distributions shown in Fig.~\ref{Figure_long_pulse_with_Lorentz}b)]. Such an inadequate overestimation of the total decay rate becomes particularly pronounced in the case of even higher values of the coupling strength as those used in Fig.~\ref{Figure_long_pulse_with_Lorentz} (see Sec.~\ref{Sec_Class_dyn} for more details). Nevertheless, it is very instructive to consider at first the simple picture associated with a Lorentzian distribution, because in this case the problem can be solved analytically giving intuitive insights into the dynamical properties of our system. By plugging the Lorentzian distribution (\ref{rho_Lorentz}) into Eq.~(\ref{Eq_a_with_Bk0}) and assuming that the cavity is initially empty, $A(0)=0$, and spins are unexcited, $B_k(0)=0$, we obtain the following Volterra equation (in the frame rotating with $\omega_p$) under the action of a rectangular microwave pulse introduced above for $t \le \tau_d$:
\begin{eqnarray}
\label{Eq_a_with_A0_Lorentz_decay}
\dot A(t)=-\kappa A(t)-\Omega^2  \int\limits_{0}^t  \! d\tau e^{-\Delta (t-\tau)}A(\tau)-\eta.
\end{eqnarray}
By differentiating Eq.~(\ref{Eq_a_with_A0_Lorentz_decay}) with respect to time, and after doing some algebra, the above equation reduces to the one for a damped harmonic oscillator driven by a time-independent external force
\begin{eqnarray}
\label{Eq_a_with_A0_Lorentz_decay_diff}
\ddot A(t)+[\Delta+\kappa] \dot A(t)+[\Omega^2+\Delta\kappa] A(t)+\eta \Delta=0.
\end{eqnarray}
The solution of Eq.~(\ref{Eq_a_with_A0_Lorentz_decay}), which is also the one of Eq.~(\ref{Eq_a_with_A0_Lorentz_decay_diff}), can be represented as $A(t)=\alpha e^{\lambda_1 t}+\beta e^{\lambda_2 t}$, so that the dynamics is characterised by two exponents, namely
\begin{eqnarray}
\label{Gam_Lorentz_charact_exponent}
\lambda_{1,2}=\left[-(\Delta+\kappa)\pm\sqrt{(\Delta-\kappa)^2-4 \Omega^2}\right]/2.
\end{eqnarray}
In the strong-coupling regime the dynamics is underdamped, the expression under the square root is negative and the system exhibits damped oscillations with the Rabi frequency 
\begin{eqnarray}
\label{Eq_Rabi_Om}
\Omega_{R}=\sqrt{4 \Omega^2-(\Delta-\kappa)^2},
\end{eqnarray}
and the decay rate of $|A(t)|^2$ is $\Gamma=\Delta+\kappa$. It is worth noting that for the case shown in Fig.~\ref{Figure_long_pulse_with_Lorentz}b), the expression (\ref{Eq_Rabi_Om}) for the Rabi frequency can be approximated as $\Omega_R \approx 2\Omega$. Finally, we obtain the following expression for the cavity amplitude for $t\le \tau_d$
\begin{eqnarray}
\label{Afunsolon}
\nonumber
&&A(t)\!=\!-\dfrac{\Delta \eta}{\Omega^2+\Delta\kappa}+\dfrac{\eta\cdot e^{-(\Delta+\kappa)t/2}}{2\Omega_{R}(\Omega^2+\Delta\kappa)}\times
\\
\\\nonumber
&&\left[2 \Omega_R \Delta \cos(\Omega_R t/2) -[\Omega_R^2-\Delta^2+\kappa^2)] \sin(\Omega_R t/2)\right].\,\,\,\,\,\,
\end{eqnarray}
The reason why $A(t)\in {\mathbb R}$ in Eq.~(\ref{Afunsolon}) is due to the fact that the Lorentzian distribution (\ref{rho_Lorentz}) is symmetric with respect to $\omega_s$, and $\omega_p=\omega_c=\omega_s$. For the same reason the $y$-component of the collective spin $J_y=0$, whereas $J_x(t)$ can easily be determined from Eq.~(\ref{Eq_a_Volt}) 
\begin{eqnarray}
\label{Jx_evol_on}
J_x(t)\!=\!\dfrac{\sum_k g_k B_k(t)}{2\Omega}\!=\!\dfrac{\dot A(t)+\kappa A(t)+\eta}{2\Omega}.
\end{eqnarray}
Indeed, by inserting the solution (\ref{Afunsolon}) into this equation we get
\begin{eqnarray}
&&J_x(t)\!=\!\dfrac{\eta\Omega}{2(\Omega^2+\Delta\kappa)}-\dfrac{\eta\Omega\cdot e^{-(\Delta+\kappa)t/2}}{2\Omega_{R}(\Omega^2+\Delta\kappa)}\times
\\\nonumber
\\
&&\left[(\Delta+\kappa) \sin(\Omega_R t/2) +\Omega_R\cos(\Omega_R t/2)\right].\,\,\,\,\,\,
\label{J_xon}
\end{eqnarray}
By differentiating Eq.~(\ref{Eq_a_Volt}) with respect to time twice, making use of Eq.~(\ref{Eq_a_with_A0_Lorentz_decay_diff}), and performing straightforward algebraic calculations, we find that $J_x(t)$ obeys also the following equation
\begin{eqnarray}
\label{Eq_a_with_Jx_Lorentz_decay_diff}
\ddot J_x(t)+\Delta \dot J_x(t)+\Omega^2 J_x(t) -\dfrac{\kappa\Omega}{2}A(t)-\dfrac{\eta\Omega}{2}=0.
\end{eqnarray}
Therefore in the case of a Lorentzian distribution the dynamics can be modelled by two coupled damped harmonic oscillators governed by Eqs.~(\ref{Eq_a_with_A0_Lorentz_decay_diff}, \ref{Eq_a_with_Jx_Lorentz_decay_diff}).

Thus, after switching on a rectangular microwave signal our system exhibits damped Rabi oscillations and it tends finally to a steady-state
\begin{eqnarray}
\label{Amp_st_Lor}
A_{st}\!=\!-\dfrac{\Delta \eta}{\Omega^2+\Delta\kappa},\,\,\,\,J_x^{st}=\dfrac{\eta\Omega}{2(\Omega^2+\Delta\kappa)},\,\,\,\,J_y^{st}=0,
\end{eqnarray}
provided that the pulse duration is long enough, i.e. $\tau_d \gg 1/(\Delta+\kappa)$. (Note that this condition is very well fulfilled in Fig.~\ref{Figure_long_pulse_with_Lorentz}.) Inserting the Lorentzian profile (\ref{rho_Lorentz}) into Eq.~(\ref{Eq_a_with_Bk0_cont_2}) from Appendix \ref{App_Decay} yields the equation for the cavity amplitude $A(t)$, which governs the decay process from the steady-state given by Eq.~(\ref{Amp_st_Lor}):
\begin{eqnarray}
\label{Eq_a_with_A0_Lorentz_decay_from_st_state}
\!\!\!\!\!\!\dot A(t)\!=\!-\kappa A(t)\!+\!\dfrac{\eta \Omega^2 \cdot e^{-\Delta t}}{\Omega^2+\kappa \Delta}
\!-\!\Omega^2  \int\limits_{0}^t  \! d\tau e^{-\Delta (t-\tau)}A(\tau),
\end{eqnarray}
where, for the sake of simplicity, the time is counted from zero as the pulse is turned off. As discussed in detail before  and also in Appendix \ref{App_Decay}, the second term in Eq.~(\ref{Eq_a_with_A0_Lorentz_decay_from_st_state}) stands for the excitation stored in the spin ensemble, which is coherently released back into the cavity, after switching off the microwave pulse. Similarly as done above, we can derive from Eq.~(\ref{Eq_a_with_A0_Lorentz_decay_from_st_state}) a damped harmonic oscillator equation, $\ddot A(t)+[\Delta+\kappa] \dot A(t)+[\Omega^2+\Delta\kappa] A(t)=0$, so that finally the damped Rabi oscillations of the cavity amplitude and the $x$-component of the collective spin to the ground state for $t\ge \tau_d$ are solved by
\begin{eqnarray}
\label{Afunsoloff}
\nonumber
\!\!\!\!\!\!\!\!\!\!&&A(t)\!=\!\dfrac{\eta\cdot e^{-(\Delta+\kappa)(t-\tau_d)/2}}{2\Omega_{R}(\Omega^2+\Delta\kappa)}\cdot\left[-2 \Omega_R \Delta \cos(\Omega_R (t-\tau_d)/2) +\right.
\\
\!\!\!\!\!\!\!\!\!\!
\\
\!\!\!\!\!\!\!\!\!\!
\nonumber
\!\!\!\!\!\!\!\!\!\!\!\!&&\left. (\Omega_R^2-\Delta^2+\kappa^2) \sin(\Omega_R (t-\tau_d)/2)\right]\!,
\\
\!\!\!\!\!\!\!\!\!\!
\nonumber
\\
\!\!\!\!\!\!\!\!\!\!
\nonumber
\!\!\!\!\!\!\!\!\!\!&&J_x(t)\!=\!\dfrac{\eta\Omega\cdot e^{-(\Delta+\kappa)(t-\tau_d)/2}}{2\Omega_{R}(\Omega^2+\Delta\kappa)}\cdot\left[(\Delta+\kappa)\sin(\Omega_R (t-\tau_d)) +\right.
\label{J_xoff}
\\
\!\!\!\!\!\!\!\!\!\!
\\
\nonumber
\!\!\!\!\!\!\!\!\!\!&&\left. \Omega_R\cos(\Omega_R (t-\tau_d)/2)\right]\!.
\end{eqnarray}
\begin{figure}
\includegraphics[angle=0,width=1.\columnwidth]{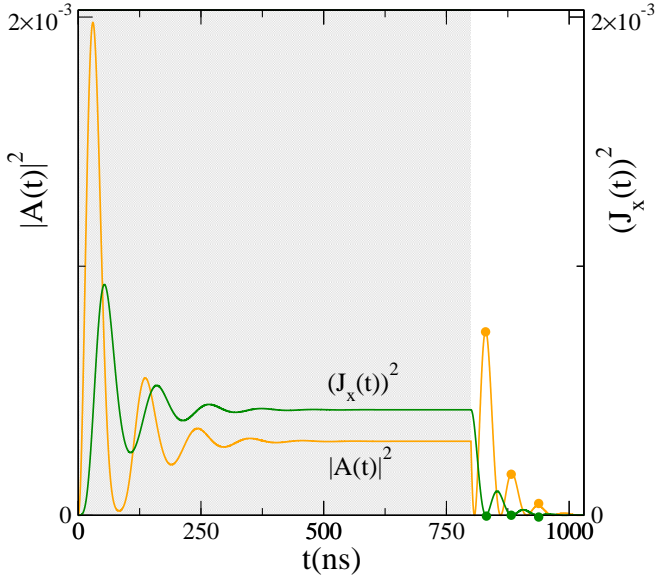}
\caption{(color online). Cavity probability amplitude $|A(t)|^2$ and the corresponding $x$-component of the collective spin $J_x^2(t)$ versus time $t$ under the action of an incident long pulse assuming a Lorentzian spin distribution, given by Eqs.~(\ref{Afunsolon},\ref{Afunsoloff}) and (\ref{J_xon},\ref{J_xoff}), respectively. $|A(t)|^2$ coincides with the orange (light gray) curve from Fig.~\ref{Figure_long_pulse_with_Lorentz}b). Symbols designate the maxima and minima of $|A(t)|^2$ and $J_x^2(t)$ during the damped Rabi oscillations. The carrier frequency matches the resonance condition, $\omega_p=\omega_c=2\pi\cdot 2.6915$\,GHz, and the frequency of Rabi oscillations, $\Omega_R=2\pi\cdot 19.2$\,MHz. Gray (white) area indicates the time interval during which the pumping signal is on (off).}
\label{Fig_A_2_Jx_2_Lor}
\end{figure}
\begin{figure}
\includegraphics[angle=0,width=1.\columnwidth]{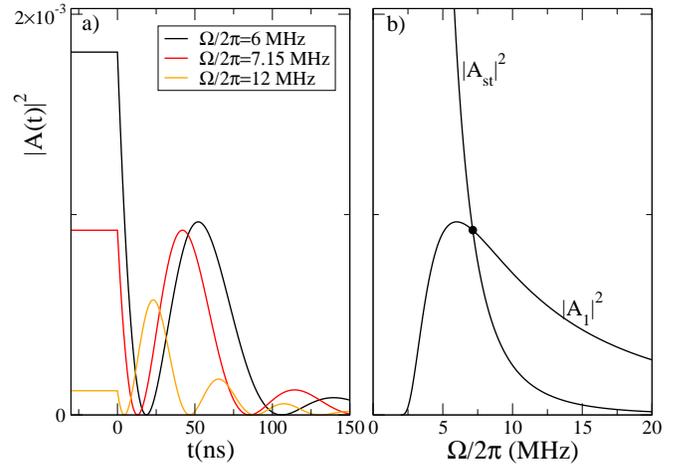}
\caption{(color online). Damped Rabi oscillations from the stationary state which the system exhibits after the action of an incident long pulse assuming a Lorentzian spin distribution.  a) Cavity probability amplitude given by Eq.~(\ref{Afunsoloff}), versus time for three different values of the coupling strengths, $\Omega/2\pi=6,7.15$ and $12$~MHz [black, red (gray), orange (light gray)]. The carrier frequency matches the resonance condition, $\omega_p=\omega_c=2\pi\cdot 2.6915$\,GHz. The lowest value for the stationary state corresponds to the highest value of $\Omega$ in accordance with Eq.~(\ref{Amp_st_Lor}). b) The amplitude of the stationary state $|A_{st}|^2$ and the amplitude of the first maximum $|A_1|^2$,  versus coupling strength $\Omega$ during the damped Rabi oscillations [see Eqs.~(\ref{Amp_st_Lor}, \ref{A1})]. Black symbol designates the intersection between these two curves at the value of coupling strength $\Omega/2\pi=7.15$~MHz, below which the overshoot effect is absent.}
\label{Figure_A0_A1_2_Lorentz}
\end{figure}
\begin{figure}
\includegraphics[angle=0,width=1.\columnwidth]{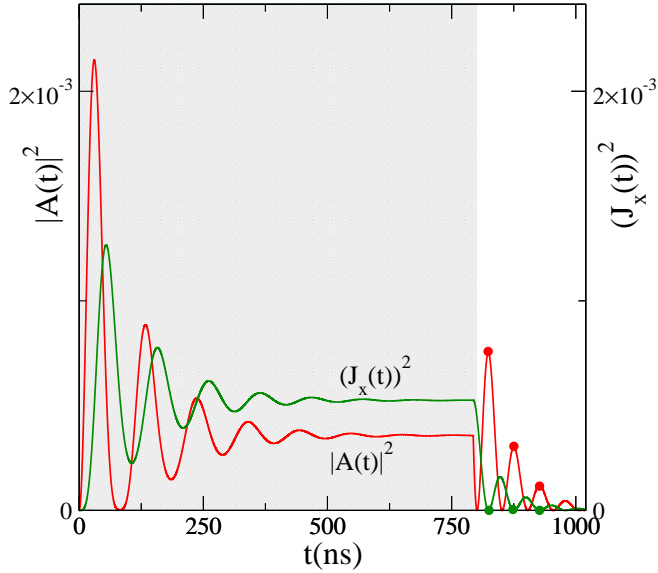}
\caption{(color online). Cavity probability amplitude, $|A(t)|^2$, and the corresponding $x$-component of the collective spin, $J_x^2(t)$, versus time $t$ under the action of an incident long pulse for the $q$-Gaussian spin distribution. $|A(t)|^2$ coincides with the red (gray) curve from Fig.~\ref{Figure_long_pulse_with_Lorentz}b). Symbols designate the maxima and minima of, respectively, $|A(t)|^2$ and $J_x^2(t)$ during the damped Rabi oscillations. The carrier frequency matches the resonance condition, $\omega_p=\omega_c=2\pi\cdot 2.6915$\,GHz, and the coupling strength $2\Omega=17.12$~MHz. The frequency of the resulting Rabi oscillations, $\Omega_R=2\pi\cdot 19.2$\,MHz. Gray (white) area indicates the time interval during which the pumping signal is on (off).}
\label{Fig_A_2_Jx_2_q_Gauss}
\end{figure}
\begin{figure}
\includegraphics[angle=0,width=1.\columnwidth]{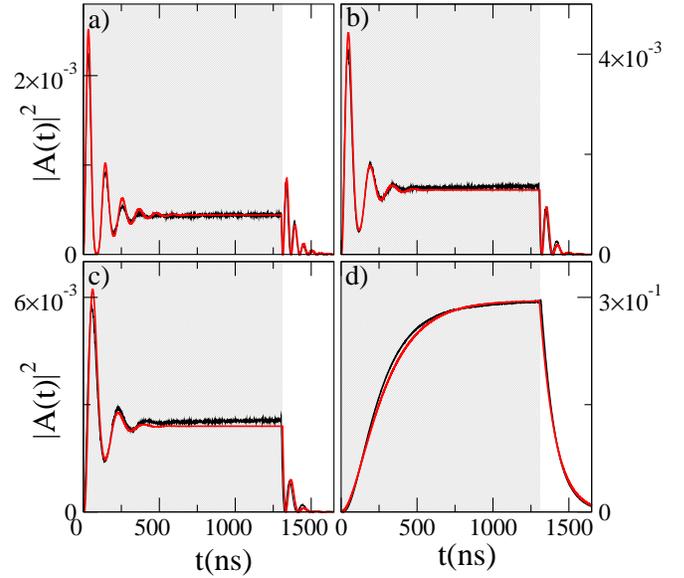}
\caption{(color online).  Cavity probability amplitude, $|A(t)|^2$, versus time $t$ under the action of an incident long pulse with the carrier frequency matching the resonance condition, $\omega_p=\omega_c=2\pi\cdot 2.6915$ GHz. The coupling strength $2\Omega$ is: a) $2\pi\cdot 15.8$\,MHz; b) $2\pi\cdot 12.0$\,MHz; c) $2\pi\cdot 10.2$\,MHz; d) $2\pi\cdot 2.12$\,MHz. Gray (white) area indicates the time interval during which a pumping signal is on (off). Red (gray) curves: results of numerical calculations. Black curves: experimental results for the cavity transmission.}
\vspace*{.3cm}
\label{fig_A_2_diff_Omega_wp_eq_wc}
\end{figure}
In Fig.~\ref{Fig_A_2_Jx_2_Lor},  $|A(t)|^2$ and $J_x^2(t)$, defined by Eqs.~(\ref{Afunsolon},\ref{Afunsoloff}) and by Eqs.~(\ref{J_xon},\ref{J_xoff}), respectively, are plotted versus time $t$. Note that this analytical solution for the cavity probability, $|A(t)|^2$, perfectly coincides with the one found numerically which is depicted in Fig.~\ref{Figure_long_pulse_with_Lorentz}b) (For that reason the analytical solution is not shown in this figure.) One sees, that the cavity and spin ensemble exchange their energies during the time evolution, so that maxima of $A^2(t)$ correspond to minima of $J_x^2(t)$ or, in other words, the energy inside the cavity is maximal at those moments of time, when the energy stored in the ensemble is entirely emitted back into the cavity.

Let us summarize the collective spin dynamics under the action of a long pulse governed by Eqs.~(\ref{J_xon},\ref{J_xoff}) in the $\omega_p$-rotating frame. Since $J_z \approx -\sqrt{N}$ is always valid, our dynamics is restricted to the vicinity of the pole of the Bloch sphere. Additionally, $J_y=0$ owing to symmetry arguments. As a rectangular microwave signal is turned on, the $x$-component, $J_x(t)$, exhibits damped Rabi oscillations starting from the ground state and tends towards a steady state, $J_x^{st}$. After the signal is switched off, $J_x(t)$, again undergoes damped  Rabi oscillations and returns to its initial state on the pole of the Bloch sphere. These spin components in the $\omega_p$-rotating frame are connected with those in the laboratory frame as follows, $J_x^{lab}(t)=J_x(t) \cos(\omega_p t)$, $J_y^{lab}(t)=J_x(t) \sin(\omega_p t)$, and $J_z^{lab}(t)=J_z(t)\approx -\sqrt{N}$. From these expressions follows that in the laboratory frame high frequency oscillations are superimposed on the damped Rabi oscillations found in the $\omega_p$-frame. Moreover the steady state in the $\omega_p$-frame is represented by a simple precession around the $z$-axis in the laboratory- frame.

We show in Fig.~\ref{Figure_A0_A1_2_Lorentz} that the first Rabi peak of the cavity amplitude after switching off the driving pulse may exceed the corresponding steady state value (overshoot effect), if the value of the coupling strength is above a certain threshold. As discussed earlier in this Section, this effect is in principle possible due to the fact that in the steady-state at constant driving nonzero energy is preliminarily stored in the spin ensemble. However, the smaller the coupling strength $\Omega$ is, the larger the value of the cavity amplitude, $|A_{st}|$, and the weaker the excitation of the spin ensemble, $|J_x^{st}|$, see Eq.~(\ref{Amp_st_Lor}). In the limiting case of $\Omega \rightarrow 0$, there is no coupling to the spin ensemble, and it remains unexcited, $J_x^{st}=0$, whereas $|A_{st}|$ acquires its maximal value, $|A_{st}|=\eta/\kappa$. The overshoot effect can be easily quantified analytically by searching for the first maximum of the decaying cavity amplitude (\ref{Afunsoloff}), which is found to be
\begin{eqnarray}
\label{A1}
A_1^2\!=\!A_{st}^2\cdot e^{-\dfrac{2(\Delta+\kappa)}{\Omega_R}\cdot \text{\normalsize{arccos}}\left[-(\Delta-\kappa)/(2\Omega)\right]}.
\end{eqnarray}
We present $A_1^2$ and $A_{st}^2$ versus coupling strength $\Omega$ in Fig.~\ref{Figure_A0_A1_2_Lorentz}b), where one can see that the overshoot effect is realized for $\Omega/2\pi>7.15$\,MHz (for the Lorentzian distribution). Note that the strong coupling regime, the hallmark of which are Rabi oscillations, terminates at $\Omega/2\pi=2$\,MHz, where $A_1=0$. At lower values of the coupling strength the oscillations do not occur and the dynamics becomes Markovian, see Sec.~\ref{Sec_Class_dyn} for more details.
\subsection{Dynamics for the $q$-Gaussian spin density distribution}
\label{Subsec_Dyn_Q_Gauss_distr}
\begin{figure}
\includegraphics[angle=0,width=1.\columnwidth]{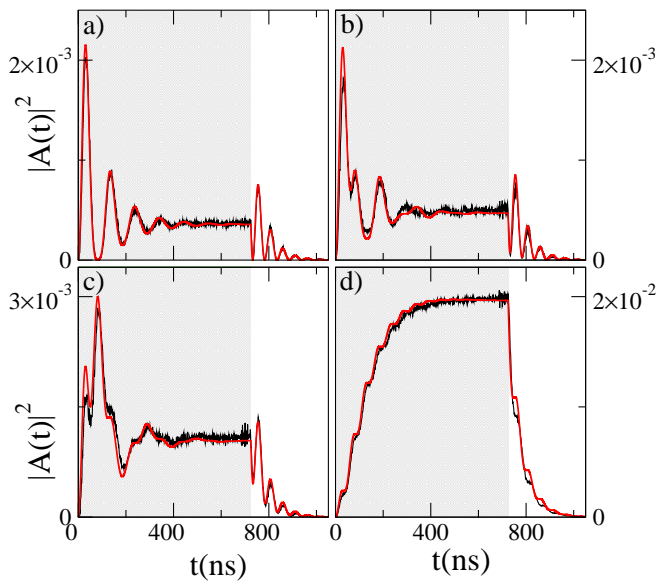}
\caption{(color online). Cavity probability amplitude, $|A(t)|^2$, versus time $t$ under the action of an incident long pulse for different values of the carrier frequency: a) $\omega_p=\omega_c$; b) $\omega_p=\omega_c\pm \Omega_R/8$; c) $\omega_p=\omega_c\pm \Omega_R/4$; d) $\omega_p=\omega_c\pm \Omega_R/2$, where $\omega_c=2\pi\cdot 2.6915$\,GHz and $\Omega_R=2\pi\cdot 19.2$\,MHz are, respectively, the cavity and Rabi frequencies. Gray (white) area indicates the time interval during which the driving signal is on (off). Red (gray) curves: results of numerical calculations for the coupling strength $2\Omega=17.12$~MHz. Black curves: experimental results for the cavity transmission.}
\label{fig_A_2_vs_t_diff_wp}
\end{figure}

After considering the case of a Lorentzian distribution for the spin density, which allows us to capture some of the important features of the dynamics, we return to the case of the $q$-Gaussian density profile to describe the dynamics accurately and to demonstrate a qualitatively new effect not existing in the framework of the Lorentzian distribution, i.e., the so-called cavity protection effect, see Sec.~\ref{Sec_cavit_prot}.  

In Fig.~\ref{Fig_A_2_Jx_2_q_Gauss} we present the coherent energy exchange between cavity and spin ensemble under the action of the long pulse, which looks rather similar to the one shown in Fig.~\ref{Fig_A_2_Jx_2_Lor} for the Lorentzian distribution. For the latter, however, our analysis predicts an overestimated decay rate with deviations that grow to an unacceptable degree for higher values of the coupling strengths as will be demonstrated in Sec.~\ref{Sec_Class_dyn}. Another signature of the non-Lorentzian line shape of our spectral spin distribution $\rho(\omega)$ is that the Rabi frequency $\Omega_R$ deviates significantly from twice the value of the coupling strength $2\Omega$. In other words, our hybrid cavity-spin system cannot be modeled as two coupled damped harmonic oscillators as in the case of a purely Lorentzian spin distribution.

In Fig.~\ref{fig_A_2_diff_Omega_wp_eq_wc} we show the dynamics under the action of a long pulse for the resonant  case, $\omega_p=\omega_c=\omega_s$, but for different values of the coupling strength $\Omega$ \cite{Omega_exp}. 
One can see in Figs.~\ref{fig_A_2_diff_Omega_wp_eq_wc}a)-d) that the steady-state value, $|A_{st}|$, increases as $\Omega$ decreases, which is in line with Eq.~(\ref{Ast_gen}). One can also see that the value of the first Rabi peak decreases with a decrease of the coupling strength. As a result, the overshoot effect fades away gradually; finally the Rabi oscillations disappear, implying that we enter the regime of Markovian dynamics. As discussed in Sec. IIIA these features are also qualitatively captured when approximating the spin density by the Lorentzian distribution.

Next, we keep the value for the coupling strength constant (staying in the strong-coupling regime) and vary the probe frequency, see Fig.~\ref{fig_A_2_vs_t_diff_wp}. The larger the mismatch from the resonance condition, $\omega_p=\omega_c=\omega_s$, the less visible the Rabi oscillations, so that finally they become completely blurred. The reason for this behavior is the following: as the probe frequency, $\omega_p$, gets increasingly detuned from the central spin frequency, $\omega_s$, the phase in the exponential function of Eq.~(\ref{Eq_a_with_Bk0}) increases at those frequencies where the contribution of $\rho(\omega)$ is non-negligible. As a consequence, during subsequent time integration the resulting integral becomes small due to the fast oscillations of the exponential function, so that the effect of strong coupling smears out. In this case the dynamics is reminiscent of the Markovian regime which occurs right at the resonance condition but for small values for $\Omega$, see Fig.~\ref{fig_A_2_diff_Omega_wp_eq_wc}d). 

We would like to emphasize, that in our numerical calculations shown in Figs.~\ref{fig_A_2_diff_Omega_wp_eq_wc}-\ref{fig_A_2_vs_t_diff_wp}, we vary only the values for the coupling strength and probe frequency, whereas all other parameters are kept the same as those in Fig.~\ref{Figure_long_pulse_with_Lorentz}a). Still, the agreement between our theoretical model and the experiment is found to be excellent.

\section{Classification of the dynamics}
\label{Sec_Class_dyn}
\begin{figure}
\includegraphics[angle=0,width=1.\columnwidth]{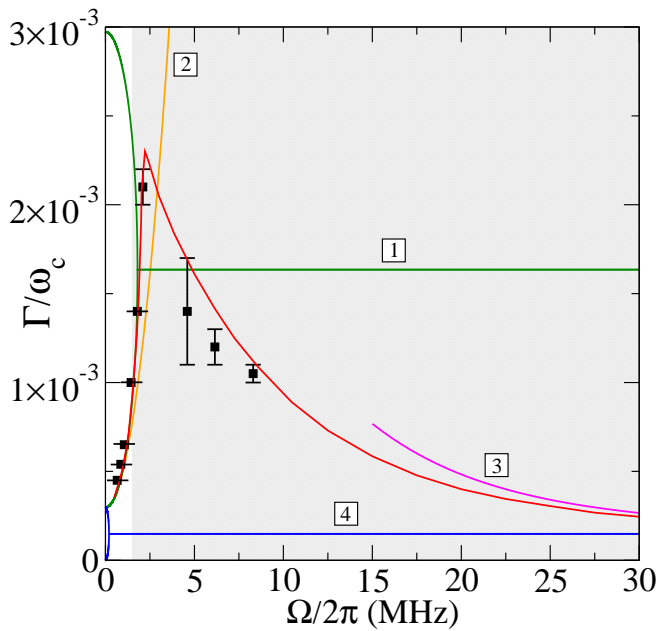}
\caption{(color online). Decay rate $\Gamma$ of the the cavity mode $|A(t)|^2$ versus coupling strength $\Omega$.
{\it Red curve}: decay rates extracted from the full numerical calculations with the $q$-Gaussian spin distribution. {\it Black symbols}: experimentally observed decay rates.  {\it Green curve (1)}: Decay rate 
under the assumption of a Lorentzian distribution of the spin density.  The overdamped regime ($\Omega/2\pi<1.8$~MHz) is characterised by two exponents given by $\Gamma=\Delta+\kappa\pm\sqrt{(\Delta-\kappa)^2-4 \Omega^2}$. The regime of underdamped oscillations ($\Omega/2\pi>1.8$~MHz) with the Rabi frequency (\ref{Eq_Rabi_Om}) has the constant decay rate, $\Gamma=\Delta+\kappa$. {\it Orange curve (2)}:  $\Gamma$ derived under Markovian approximation, $\Gamma=2[\kappa+\pi\Omega^2\rho(\omega_s)]$. {\it Magenta curve (3)}: an estimate for $\Gamma$ within the strong coupling regime with a well-resolved Rabi splitting in the limit of $\Omega\rightarrow\infty$, $\Gamma=\kappa+\pi\Omega^2\rho(\omega_c\pm\Omega)$. {\it Blue curve (4)}: the decay rate in the absence of dephasing. For $\Omega/2\pi<0.2$~MHz the overdamped regime is characterised by two exponents $\Gamma=\kappa\pm\sqrt{\kappa^2-4 \Omega^2}$. In the opposite case, $\Omega/2\pi>0.2$~MHz, the regime of underdamped Rabi oscillations takes place with the Rabi frequency $\sqrt{4 \Omega^2-\kappa^2}$ and the constant decay rate  $\Gamma=\kappa$. {\it White region}: Markovian dynamics. {\it Gray region}: non-Markovian dynamics.}
\label{Figure_Gamma_vs_Omega}
\end{figure}

To clarify the role played by the non-Lorentzian inhomogeneous broadening, we classify the dynamics by calculating and measuring the total decay rate $\Gamma$ of the cavity amplitude squared, $|A(t)|^2$, from its steady state value  for different coupling strengths $\Omega$. For the sake of simplicity, we focus on the resonant case, $\omega_p=\omega_c=\omega_s$, only. It should be stressed that the total decay rate $\Gamma$ is independent of the initial conditions (see also \cite{Nature2014}), so that we can start from simpler initial conditions corresponding to the case when only a single photon is populating the cavity and the spin ensemble is in the ground state, $|1,G \rangle=a^{\dagger}(t=0)|0 \rangle$ ($|0 \rangle$ corresponds to the vacuum state). In this case it is possible to get a relatively simple form for the Laplace transform of the Volterra equation and to considerably speed up the calculations, see Appendix~\ref{App_Lapl_Transf}. One can prove that the Volterra equation (\ref{Eq_a_with_Bk0}) is indeed the governing equation for $A(t)$ also in this case with the initial condition, $A(t=0)=1$ and $B_k(t=0)=0$, by virtue of the following arguments. Acting with the Heisenberg operator equations on the bra- and ket-vectors $\langle 0|$ and $a^{\dagger}(t=0)|0 \rangle$, respectively, it can be shown that the corresponding equations for the expectation values coincide with Eqs.~(\ref{Eq_a_Volt},\ref{Eq_bk_Volt}) from Sec.~\ref{Sec_Theory}. The only formal difference now is that the amplitudes $A(t)$ and $B_k(t)$ are given as $A(t)\equiv \langle 0| a(t)a^{\dagger}(t=0)|0 \rangle$ and $B_k(t)\equiv \langle 0|\sigma_k^-(t) a^{\dagger}(t=0)|0 \rangle$, respectively. Thus the variable $A(t)$ describes the probability amplitude for a photon to be in the cavity at time $t$, if it was there initially, $A(t=0)\equiv \langle 0| a(t=0)a^{\dagger}(t=0)|0 \rangle=\langle 1,G |1,G \rangle=1$. 

The results are presented in Fig.~\ref{Figure_Gamma_vs_Omega}, where we show that the decay rate varies surprisingly strongly and in a non-monotonous fashion with $\Omega$ covering a range of almost one order of magnitude (see the red curve on this figure). Before going to further details, let us analyze at first how the decay rate $\Gamma$ behaves as a function of the coupling strength under different simplifying assumptions. 

For the case of a Lorentzian distribution for the spin density, the decay process is characterized by two exponents given by Eq.~(\ref{Gam_Lorentz_charact_exponent}). If $4 \Omega^2>(\Delta-\kappa)^2$, then the Rabi oscillations are underdamped and the total decay rate reduces to $\Gamma=\Delta+\kappa$. In the opposite case, we are dealing with a pure exponential decay without oscillations (overdamped regime) with $\Gamma=\Delta+\kappa\pm\sqrt{(\Delta-\kappa)^2-4 \Omega^2}$. Thus, the Lorentzian distribution gives rise to qualitatively different behaviour for the decay process as compared to the $q$-Gaussian one, since $\Gamma$ remains constant in the whole range of $\Omega$ within the strong-coupling regime. However, as is unambiguously seen in Fig.~\ref{Figure_Gamma_vs_Omega}, the  non-monotonic behavior obtained in the framework of the $q$-Gaussian spin density distribution is supported by our experimental data thereby confirming our initial assumption for the shape of this distribution.

In the absence of inhomogeneous broadening, when the spin density function is written as $\rho(\omega)=\delta(\omega-\omega_s)$, the expressions for the decay rate are obtained from those for a Lorentzian distribution by setting its width to zero, $\Delta=0$. Thus, in the regime of underdamped oscillations we get, $\Gamma=\kappa$, whereas in the overdamped regime, $\Gamma=\kappa\pm\sqrt{\kappa^2-4 \Omega^2}$. Correspondingly, the blue lines in Fig.~\ref{Figure_Gamma_vs_Omega} determine the lowest border for possible decay rates reached in our system, because the values for $\Gamma$ in the presence of inhomogeneous broadening should always be larger than the corresponding ones in the case when it is absent. It is seen from Fig.~\ref{Figure_Gamma_vs_Omega} that this condition is indeed always fulfilled.


Next, we apply the so-called Markov approximation in Eq.~(\ref{Eq_a_with_Bk0}) with respect to the cavity amplitude $A(t)$ which implies that the memory effects caused by a feedback from the NV ensemble onto the cavity are disregarded. Specifically, we shift the initial time of integration on the r.h.s.\,of  Eq.~(\ref{Eq_a_with_Bk0})  to $-\infty$, put $A(\tau) \approx A(t)$, and make use of the Sokhotski-Plemelj theorem (\ref{Sokhotski_theorem}) in the limit of $\gamma\rightarrow 0$, when performing the integration with respect to $\omega$. Under all these assumptions the third term on the r.h.s. of Eq.~(\ref{Eq_a_with_Bk0}) reduces to ($\omega_p=\omega_s$)
\begin{eqnarray}
\nonumber
&&-\Omega^2 \int_0^{\infty} d\omega \rho(\omega) \int_{0}^t d\tau
e^{-i(\omega-\omega_s-i\gamma)(t-\tau)}A(\tau) \approx 
\\
&&i \Omega^2 A(t) \int_0^{\infty}\dfrac{d\omega \rho(\omega)}{\omega-\omega_s-i \gamma}=-\pi\Omega^2\rho(\omega_s)\cdot A(t).
\label{Eq_Mark_Volt}
\end{eqnarray}
Note, that the principal value does not appear in the above equation because $\rho(\omega)/(\omega-\omega_s)$, is an antisymmetric function with respect to the singular point, $\omega=\omega_s$. In the simplest case when there is no driving and all spins are initially in the ground state, the Volterra equation (\ref{Eq_a_with_Bk0}) reduces to $\dot A(t)\!=-[\kappa+\pi\Omega^2\rho(\omega_s)]\cdot A(t)$. Therefore, the Markov approximation leads to a pure exponential decay with the decay rate, $\Gamma=2[\kappa+\pi\Omega^2\rho(\omega_s)]$. The spin ensemble density thus gives rise to a significant enhancement of the cavity decay rate as compared to the one for a bare cavity, $\Gamma=2\kappa$. Remarkably, this effect has a direct analogy to the Purcell enhancement of the spontaneous emission rate of a single emitter inside a cavity \cite{Purcell1946} which appears due to the increase of the local density of photonic states at the emitter position as compared to the vacuum case. The Markov approximation, however, loses its validity at fairly low coupling strengths, starting to deviate from the real values of $\Gamma$ already at $\Omega/2\pi \approx 1.5$\,MHz (see Fig.~\ref{Figure_Gamma_vs_Omega}). The hallmark of non-Markovianity of the resulting dynamics are Rabi oscillations setting in at higher values of $\Omega$.

In a next step we put forward an analytical estimate for the decay rate in the limit of very strong coupling ($\Omega \rightarrow \infty$) employing the Laplace transform of our Volterra equation summarized in Appendix \ref{App_Lapl_Transf}. For that purpose we use recently developed concepts for another cavity QED problem dealing with non-Markovian quantum dynamics of a single emitter inside an open multimode cavity \cite{KLRT14}. The key insight from that study is that the dominant frequency components contributing to the dynamics of $A(t)$ are those which are resonant in its Laplace transform, $U(\omega)$, given by Eq.~(\ref{Eq_Phis_34}). For such resonances to occur we find the following requirement on the nonlinear Lamb shift (\ref{Eq_Lamb_shift}), $\omega_r-\omega_c= \Omega^2\delta( \omega_r)$. In the limit of sufficiently large values of the coupling strength the Laplace transform, $U(\omega)$, has a well-resolved double-peak structure with two resonance frequencies given approximately by, $\omega_r \approx\omega_c\pm\Omega$. Furthermore, $A(t)$ essentially displays damped Rabi oscillations of the form, $A(t)\sim \cos(\Omega t)\cdot e^{-[\kappa+\pi\Omega^2\rho(\omega_c\pm\Omega)]t/2}$, due to the Fourier transforms of the two curves in $U(\omega)$ centered at these two resonance frequencies. One can see in Fig.~\ref{Figure_Gamma_vs_Omega} that such an estimate for the decay rate, $\Gamma=\kappa+\pi\Omega^2\rho(\omega_c\pm\Omega)$, works rather well if $\Omega/2\pi \ge 25$\,MHz. Thus, in contrast to the Markovian dynamics, the relevant frequencies which contribute to the value of the decay rate are those associated with two resonant peaks in $U(\omega)$. Remarkably, a pair of poles in the complex plane occurring for $\Omega/2\pi \ge 25$\,MHz do not spoil this asymptotic behavior, see Appendix \ref{App_Lapl_Transf}. Note that our expression for the decay rate in the limit of  $\Omega \rightarrow \infty$ coincides with the one obtained in \cite{Diniz2011}, where the behavior of poles of the stationary transmission has been analyzed.

\subsection*{Cavity protection effect}
\label{Sec_cavit_prot}
It follows from the above analysis that for spectral distributions $\rho(\omega)$ whose tails fall off faster than  
$1/\omega^2$, an increasing coupling strength inevitably leads to a reduction of the decay rate $\Gamma$, so that the system will finally be protected against decoherence, a phenomenon referred to as ``cavity protection effect'' \cite{Kurucz2011,Diniz2011}. It is not hard to see that our $q$-Gaussian satisfies such a requirement, whereas a Lorentzian spin distribution does not. As a consequence, the latter does not protect the cavity against decoherence, featuring a constant decay rate in the strong coupling regime (see green line in Fig.~\ref{Figure_Gamma_vs_Omega}). In contrast, our numerical analysis for the $q$-Gaussian shows that for a collective coupling strength of $\Omega/2\pi \sim 25$\,MHz, the decay rate induced is already suppressed below $8\%$ of its maximal value at $\Omega/2\pi \sim 2.25$\,MHz. It is interesting to note, that the minimal possible value for the decay rate reached in the limit of large $\Omega$ is $\kappa$ as the decay rate for a bare cavity without diamond is $2\kappa$. This can be explained by the fact that due to the strong coupling between the spin ensemble and the cavity, the excitation is trapped by $50\%$ within the spin ensemble which has a negligible direct decay rate during the course of our experiment.

Physically, the ``cavity protection effect" can  be understood as follows: In the presence of inhomogeneous spin broadening, the polariton states, defined as superpositions of the cavity mode with the superradiant (bright) spin-wave modes, become coupled to the sub-radiant (dark) spin-wave modes \cite{Kurucz2011}. This coupling acts as the main source of decoherence, leading to a strong damping of the polariton modes. However, for strong enough coupling strength, the Rabi-splitting of the polariton peaks opens up a gap for the super-radiant polaritons.  If the spectral profile of the inhomogeneous spin distribution decays sufficiently fast for increasing gap size, an energetic decoupling of the super-radiant polaritons from the sub-radiant spin-wave modes occurs, leading to a suppressed damping of the polaritons and to a corresponding decrease of their peak linewidth.
\begin{figure}
\includegraphics[angle=0,width=1.\columnwidth]{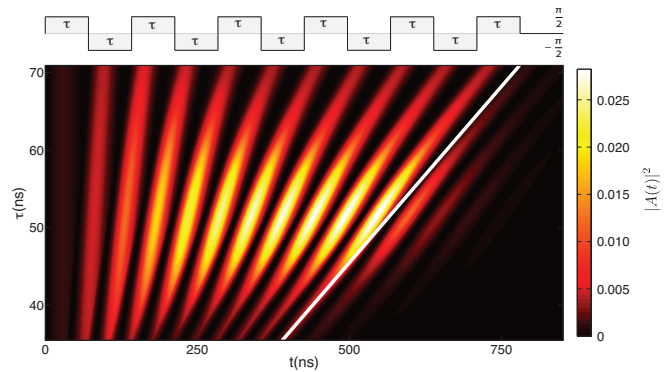}
\caption{(color online). Cavity probability amplitude $|A(t)|^2$ under the action of eleven successive rectangular microwave pulses with carrier frequency $\omega_p=\omega_c=\omega_s=2\pi\cdot 2.6915$\,GHz, phase-switched by $\pi$, as a function of time and pulse duration $\tau$. The white line indicates the corresponding moment of times, $11 \tau$, at which the driving signal is switched off. The coupling strength $2\Omega/2\pi=17.12$\,MHz.}
\label{fig_6_pulses_2D}
\end{figure}
\begin{figure}
\vspace*{0.84cm}
\includegraphics[angle=0,width=1.\columnwidth]{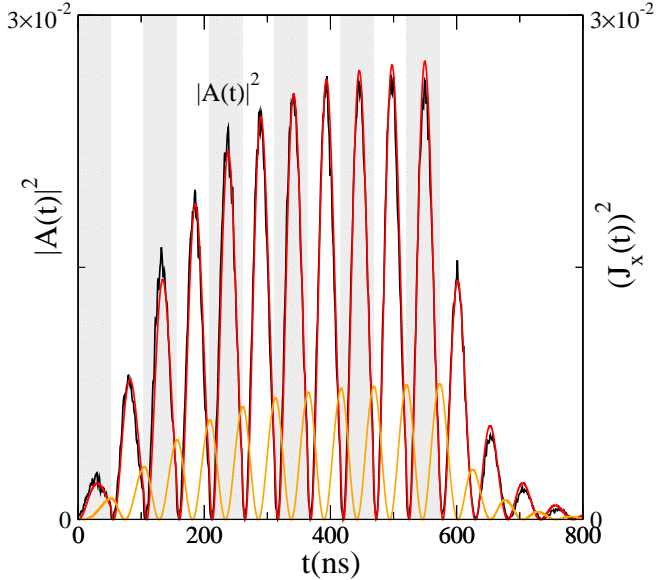}
\caption{(color online). Resonant dynamics under the action of eleven successive rectangular microwave pulses (horizontal cut of Fig.~\ref{fig_6_pulses_2D} at $\tau=2\pi/\Omega_R=52$\,ns). This specific driving corresponds to the largest enhancement of both the cavity amplitude $|A(t)|^2$ and the $x$-component of the collective spin $J_x^2(t)$ which coherently exchange the energy during course of time. Red (gray) curve: results of numerical calculations for $|A(t)|^2$. Black curve: $|A(t)|^2$ measured in the experiment. Orange (light gray) curve: results of numerical calculations for $J_x^2(t)$. The alternating gray and white vertical bars designate the pulses sketched at the top of Fig.~\ref{fig_6_pulses_2D}. The last white area corresponds to the damped dynamics when the driving signal is switched off.}
\label{Fig_A_2_Jx_2_q_Gauss_6_pulses}
\end{figure}
\section{Coherently driven spin ensembles}
\label{Sec_period_drive}

In a next step we address an important question arising in the context of possible realizations of coherent-control schemes, which is how to reach high excitation levels in the spin ensemble with a driving signal that has only limited power to avoid heating up the hybrid quantum device. We have seen in Sec.~\ref{Subsec_Dyn_Lorentz_distr} that the assumption of a Lorentzian distribution for the spin density leads to a simplified picture reducing the dynamics to the one of two coupled damped harmonic oscillators, where one of them stands for the cavity and the other for the spin ensemble. Furthermore, the expectation value of the collective spin operator can formally be excluded, so that we end up with a single equation for the cavity amplitude which has the same form as the equation for a damped and driven harmonic oscillator. Therefore, if our system is subjected to a periodic driving force, a resonance is expected to occur when the driving frequency is equal to the characteristic frequency of the system. Based on this reduced model, we conjecture that coherent cavity oscillations, and as a consequence, spin ensemble oscillations with a large amplitude can also be achieved for the $q$-Gaussian spin distribution. Also in this case the system needs to be driven periodically, so that the period of $\eta(t)$ matches the resonance condition given by the Rabi period, $T_R=2\pi/\Omega_R$. 

By pumping the cavity by a sequence of rectangular pulses with a carrier frequency $\omega_p=\omega_c=\omega_s$, phase-switched by $\pi$, we indeed reveal a strongly resonant structure of $|A(t)|^2$ as a function of pulse duration $\tau$ and time $t$, see Fig.~\ref{fig_6_pulses_2D}. The corresponding increase of $|A(t)|^2$ can reach two orders of magnitude as compared to the case when the system is driven by a long rectangular pulse [see Fig.~\ref{Figure_long_pulse_with_Lorentz}a)], provided that the resonance condition is met, $\tau=2\pi/\Omega_R$ (see  Fig.~\ref{Fig_A_2_Jx_2_q_Gauss_6_pulses}). Note that the net power injected into the cavity, when applying a long rectangular pulse or a sequence of rectangular pulses phase-switched by $\pi$, is exactly the same as we are just periodically changing the sign of the amplitude. Also in both cases the cavity and spin ensemble coherently exchange their energy, so that the cavity amplitude $|A(t)|^2$ oscillates in antiphase with respect to the spin ensemble component $J_x^2(t)$.

\begin{figure}
\includegraphics[angle=0,width=1.\columnwidth]{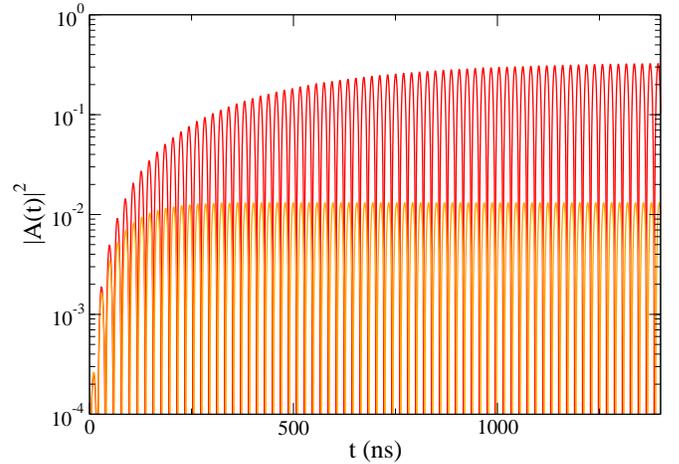}
\caption{(color online). Resonant dynamics under the action of seventy successive rectangular microwave pulses ($\omega_p=\omega_c=\omega_s$) for a pulse duration $\tau=2 \pi/\Omega_R=19.5$\,ns. Red (gray) curve: numerical results for the $q$-Gaussian spin distribution. The coupling strength is chosen to be $\Omega/2\pi=25$\,MHz. In this case the value for the total decay rate $\Gamma$ (see Fig.~\ref{Figure_Gamma_vs_Omega}) is $3.7$ times smaller than that for $\Omega/2\pi=8.56$\,MHz used so far in Figs.~\ref{fig_6_pulses_2D}, \ref{Fig_A_2_Jx_2_q_Gauss_6_pulses}. Orange (light gray) curve: corresponding numerical results for the Lorentzian spin distribution. }
\label{Fig_A_2_q_Gauss_vs_Lorentz_6_pulses}
\vspace*{0.5cm}
\end{figure}

In Fig.~\ref{Fig_A_2_q_Gauss_vs_Lorentz_6_pulses} we present results for such a resonant driving both for a $q$-Gaussian and for a Lorentzian profile of the spectral distribution for the spin density. We take the value of the coupling strength, $\Omega/2\pi=25$\,MHz, for which the decoherence effect caused by the $q$-Gaussian form of the inhomogeneous broadening is strongly suppressed. Indeed, the resulting total decay rate shown in Fig.~\ref{Figure_Gamma_vs_Omega} for this value of $\Omega$ is $3.7$ times smaller than that for $\Omega/2\pi=8.56$\,MHz used so far in Figs.~\ref{fig_6_pulses_2D}, \ref{Fig_A_2_Jx_2_q_Gauss_6_pulses} and 5.4 times smaller than the total decay rate predicted in the framework of the Lorentzian distribution. For this situation we see that the giant oscillations of the cavity probability amplitude, $|A(t)|^2$, induced by the resonant driving is a factor of 20 larger than what would be predicted for by a Lorentzian functional profile. This clear signature of the ``cavity-protection effect'' paves the way for the realization of sophisticated coherent-control schemes in the strong-coupling regime of QED.

\begin{figure}
\includegraphics[angle=0,width=1.\columnwidth]{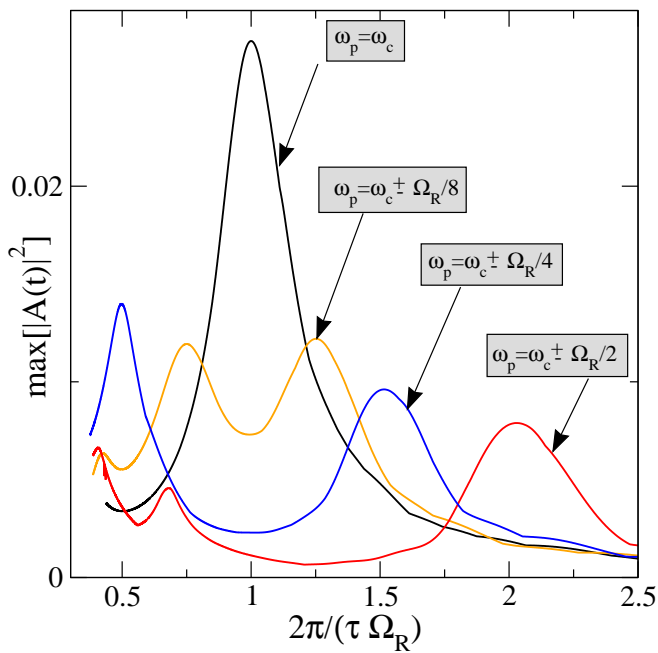}
\caption{(color online). The maximal value of the cavity probability amplitude $|A(t)|^2$, $\max[|A(t)|^2]$, reached during coherent oscillations to which the system sets in under the action of incident rectangular pulses of duration $\tau$ that are phase-switched by $\pi$. We consider four different values for the carrier frequency of our periodic driving signal: $\omega_p=\omega_c$; $\omega_p=\omega_c\pm \Omega_R/8$; $\omega_p=\omega_c\pm \Omega_R/4$; $\omega_p=\omega_c\pm \Omega_R/2$, where $\omega_c=2\pi\cdot 2.6915$\,GHz and $\Omega_R=2\pi\cdot 19.2$\,MHz are, respectively, the cavity and Rabi frequencies.}
\label{fig_A2_max_in_units_Omega_fin}
\end{figure}

In a further study we take the probe frequency out of resonance with the cavity  $\omega_p \ne \omega_c$. (The condition $\omega_c=\omega_s$, however, always holds.) In Fig.~\ref{fig_A2_max_in_units_Omega_fin} we present the maximal value of the cavity amplitude, $\max[|A(t)|^2]$, reached during coherent oscillations to which the system sets in under the action of incident rectangular pulses of duration $\tau$ that are phase-switched by $\pi$. We deduce from this figure that the cavity amplitude experiences maximal growth at the resonance condition, $\omega_p=\omega_c=\omega_s$. It is worth noting that for the off-resonant cases ($\omega_p \ne \omega_c$) the right peak of $\max[|A(t)|^2]$ appears exactly at such values of $\pi/\tau$ which correspond to the mismatching value of the probe frequency from the resonant case ($\omega_p=\omega_c$). A similar tendency is also seen for the left peak for not too high values of the mismatch from the resonance condition.

\section{Conclusions}
We have studied in detail the dynamics of an inhomogeneously broadened spin ensemble interacting with a single cavity mode. For that purpose we numerically solved the Volterra integral equation for the cavity amplitude which explicitly contains the spin distribution function describing the inhomogeneous broadening of the spin ensemble. By assuming a Lorentzian functional profile for the spin density, we solved the problem analytically. This analytical solution provides an  intuitive understanding of some important features of the resulting spin-cavity dynamics, such as an overshoot effect resulting from the constructive interference between the energy stored in the spin ensemble and in the cavity. Several features of the temporal dynamics in the strong coupling regime are, however, specifically due to the $q$-Gaussian spectral spin density which we find to be realized in our experiment. In particular, the non-Lorentzian functional profile of the spin distribution allows us to observe as well as to accurately describe a phenomenon known as ``cavity protection effect" \cite{Kurucz2011,Diniz2011} for large values of the coupling strength. This effect results in a complete suppression of the decoherence induced by inhomogeneous broadening in the strong-coupling regime. To highlight the potential of this effect for the implementation of coherent-control schemes, we reveal how an appropriately chosen pulse sequence can excite giant coherent oscillations between the cavity and the spin ensemble. We classify the dynamics as a function of the coupling strength and the probe frequency covering both Markovian and non-Markovian regimes. 


%
\section*{Acknowledgements}

We would like to thank R. Ams\"uss, B. Hartl, F. Mintert, T. N\"obauer, P. Rabl, J. Schmiedmayer and A. Valookaran for helpful discussions.  D.O.K. and S.R acknowledge funding by the FWF through Project No. F49-P10 (SFB NextLite).  The experimental effort has been supported by the TOP grant of TU Vienna. S.P. acknowledges support by the Austrian Science Fund (FWF) in the framework of the Doctoral School ``Building Solids for Function" (Project W1243).

\appendix

%
\section{Direct time integration of the Volterra equation}
\label{App_time_integr_Volt_Eq}

Although  Eq.~(\ref{Volt_eq}) has a relatively simple form, it is a challenging task to solve it numerically. There are two reasons for that:  First, in order to calculate the cavity amplitude at time $t$, one should know  the amplitude $A(\tau)$ at all previous instants, $\tau<t$ (memory effect). Second, an integration with respect to the frequency in the kernel function ${\cal K}(t-\tau)$ has to be performed for each $t$ and $\tau<t$ [see Eq.~(\ref{Volt_eq_K})]. The smallest possible time scale in our problem is given by $T=2\pi/\omega_p\sim 0.4\,$ns. To achieve a very good accuracy of the calculations, we solve the equation on a temporal mesh with uniform spacing, choosing a time step $dt \sim 0.05\,$ns (see e.g. \cite{NumRec} for more details about the method). The direct discretization of ${\cal K}(t-\tau)$ on the time interval of the order of $\mu$s (typical time of measurements) leads to a high-dimensional matrix (of a size typically exceeding $10^4 \times 10^4$), which, together with the integration with respect to frequency, makes the problem computationally intractable by way of a direct numerical solution. 

To overcome this problem and to speed up the calculations drastically, we divide the whole time integration into many successive subintervals, $T_{ n}\leq t \leq T_{n+1}$, with $n=1,2,...$. Such a time division can, in principle, be implemented arbitrarily but we choose it to be adapted to our experimental realization. Specifically, for a sequence 
of rectangular pulses with phase inversion, the driving amplitude is unchanged within each subinterval, so that $\eta(t)$ is written as $\eta_n=(-1)^{n+1}\cdot \eta$, where $n=1,2,3,...$. Thus, in order to proceed with the integration on the $n$-th time interval, which starts from the initial value $A^{n}(T_n)$, we have to provide the result of integration obtained in the previous step, $A^{(n-1)}(T_n)$. The recurrence relation (time runs within $T_{ n}\leq t \leq T_{n+1}$ for $n=1,2,3,...$) then reads
\begin{eqnarray}
\label{Volt_eq_itter}
A^{(n)}(t)=\int\limits_{T_n}^t d\tau {\cal K}(t-\tau) A^{(n)}(\tau)+{\cal F}^{(n)}(t),
\end{eqnarray}
where the kernel function ${\cal K}(t-\tau)$ is defined by Eq.~(\ref{Volt_eq_K}) and
\begin{eqnarray}
\nonumber
&&{\cal F}^{(n)}(t)=e^{-i(\omega_c-\omega_p-i\kappa)(t-T_n)} \left\{A^{(n-1)}(T_n)+\right.
\\\nonumber
&&\left.\Omega^2 \bigintsss_0^{\infty}\!\!\! d\omega\,
\dfrac{\rho(\omega) \left[e^{-i (\omega-\omega_c+i \kappa)(t-T_n)}-1\right]
}{i (\omega-\omega_c+i\kappa)}\cdot {\cal I}_n(\omega)\right\}-
\\
&&\dfrac{i{\cal\eta}_n}{\omega_c-\omega_p-i\kappa}\cdot\left[1-e^{-i(\omega_c-\omega_p-i\kappa)(t-T_n)} \right]
\end{eqnarray}
Note also that the memory on previous events enters not only through the amplitude $A^{(n-1)}(T_n)$ but also through the function  
\begin{eqnarray}
&&{\cal I}_n(\omega)=e^{-i(\omega-\omega_p)(T_n-T_{n-1})}{\cal I}_{n-1}(\omega)+
\\\nonumber
&&\int\limits_{T_{n-1}}^{T_n} d\tau e^{-i(\omega-\omega_p)(T_n-\tau)}A^{(n-1)}(\tau).
\end{eqnarray}
The initial conditions at $t=T_1=0$ are defined as $A(T_1)=0$ and $ {\cal I}_1(\omega)=0$ if the cavity is empty and spins are in the ground state.

The above technique allows us to solve Eq.~(\ref{Volt_eq}) accurately while being very efficient in terms of computational time. We have tested the accuracy of our numerical results by varying the discretization both in time and frequency in a wide range obtaining excellent agreement with the experimental results, thereby confirming the accuracy of our method.

\section{Laplace transform of the Volterra equation}
\label{App_Lapl_Transf}
\begin{figure}
\includegraphics[angle=0,width=.5\columnwidth]{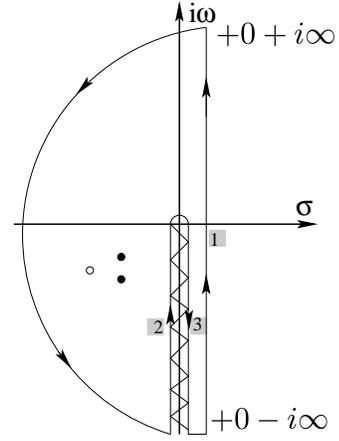}
\caption{Contour completion in the complex plane $s=\sigma+i \omega$ for the calculation of the inverse Laplace transform. Those contours which give nonzero contribution are designated by numbers. Empty circle: the pole which appears in the regime of weak coupling for $\Omega/2\pi \le 1.7$\,MHz (Markovian dynamics). Two filled circles: two poles which show up in the strong coupling regime for $\Omega/2\pi \ge 25$\,MHz. The zig-zag line corresponds to the branch cut along the negative part of the imaginary axis.}
\label{fig1_contour}
\vspace*{0.cm}
\end{figure}

In order to speed up the calculations of the decay rate for different values of the coupling strengths, 
$\Omega$, and to derive an analytical expression for it in the limit of $\Omega\rightarrow \infty$, we perform a Laplace transformation of the Volterra equation
\begin{equation}
\dot A(t)\!=\!-\!\kappa A(t)\!-\!\Omega^2 \!\int_0^{\infty}\! d\omega \rho(\omega)\! \int\limits_{0}^t \! d\tau
e^{-i(\omega-\omega_p)(t-\tau)}A(\tau),
\label{Eq_Volt_App}
\end{equation}
assuming that the decay process starts from the most simple initial condition, $A(t=0)=1$, when the cavity is fed with a single photon and the spin ensemble is in the ground state. For the sake of simplicity we consider the resonant case only, $\omega_p=\omega_c=\omega_s$. To carry out the Laplace transformation we multiply Eq.~(\ref{Eq_Volt_App}) by $e^{-st}$ and integrate both sides of the equation with respect to time from $0$ to $\infty$ (see e.g. \cite{Riley} for more details). Here $s=\sigma+i\omega$ is the complex variable so that we reformulate our problem by solving it in the complex plane of $s$. After straightforward calculations, the algebraic equation for the Laplace transform,
$\tilde A(s)=\int_0^\infty dt \,e^{-st} A(t)$, is derived which is solved by
\begin{equation}
\tilde A(s)=\dfrac{1}{s+\kappa+\Omega^2 \int_0^{\infty}
\dfrac{d\omega\rho(\omega)}{s+i(\omega-\omega_c)}}.
\end{equation}
By performing the inverse Laplace transformation, $A(t)=\dfrac{1}{2\pi i}\int_{\sigma-i \infty}^{\sigma+i \infty}ds\, e^{st} \tilde A(s)$, we obtain the
following formal solution for the cavity amplitude $A(t)$
\begin{eqnarray}
\label{Inverse_LT}
A(t)\!=\!\dfrac{e^{i\omega_c t}}{2\pi i}
\int_{\sigma-i \infty}^{\sigma+i \infty} \dfrac{e^{st}ds}{s+\kappa+i\omega_c+\Omega^2 \int_0^{\infty}
\dfrac{d\omega\rho(\omega)}{s+i\omega}},\,\,\,\,
\end{eqnarray}
where $\sigma>0$ is chosen such that the real parts of all singularities of $\tilde A(s)$ are smaller than $\sigma$. It can be proved that the integral in the denominator of Eq.~(\ref{Inverse_LT}) has a jump when passing across the negative part of the imaginary axis leading to the branch cut in the complex plane of $s$ (see Fig.~\ref{fig1_contour} and also \cite{KLRT14}). 

By setting the denominator of the integrand in Eq.~(\ref{Inverse_LT})  to zero, we derive the following equations for a simple pole, $s_j=\sigma_j+i\omega_j$,
\begin{eqnarray}
\label{Eq_poles_1}
\sigma_j&=&-\dfrac{\kappa}{1+\Omega^2\int_0^{\infty}\dfrac{d\omega\rho(\omega)}{\sigma_j^2+(\omega_j+\omega)^2}},
\\
\omega_j&=&-\omega_c+\Omega^2\int_0^{\infty}\dfrac{d\omega\rho(\omega)(\omega_j+\omega)}{\sigma_j^2+(\omega_j+\omega)^2}.
\label{Eq_poles_2}
\end{eqnarray}
It turns out that a single solution of Eqs.~(\ref{Eq_poles_1}-\ref{Eq_poles_2}) exists within the weak coupling regime in a rather narrow interval of the coupling strengths, $\Omega/2\pi \le 1.7$\,MHz. (It is seen that in the limit of $\Omega\rightarrow 0$, Eqs.~(\ref{Eq_poles_1}-\ref{Eq_poles_2}) are solved by $\sigma_j\sim-\kappa$ and $\omega_j=-\omega_c$, respectively.) We also find a pair of poles with $\sigma_1=\sigma_2<0$, $|\sigma_{1,2}|\ll \kappa$ and $\omega_{1,2}=-\omega_c\pm \epsilon$ in the strong coupling regime for large values of the coupling strength starting from $\Omega/2\pi \approx 25$\,MHz. Note that both $|\sigma_{1,2}|$ and $\epsilon$ grow with increasing $\Omega$.


Next, we apply Cauchy's theorem to a closed contour to evaluate the original integral Eq.~(\ref{Inverse_LT}) taking into account that only a few paths of those shown in Fig.~\ref{fig1_contour} contribute. Finally, we end up with the following expression for the cavity amplitude $A(t)$
\begin{eqnarray}
A(t)=e^{i\omega_c t}\left\{\Omega^2\int_{0}^{\infty} d\omega e^{-i\omega t} U(\omega)+\sum_j R_j\right\},\,\,\,\,\,
\end{eqnarray}
where
\begin{eqnarray}
\label{Eq_Phis_34}
&&U(\omega)=\lim_{\sigma\rightarrow 0^{+}}
\\
\nonumber
&&\left\{\dfrac{\rho(\omega)}
{\left(\omega\!-\!\omega_c\!-\!\Omega^2 \delta(\omega)+\!i\kappa\right)^2\!+\!(\pi\Omega^2\rho(\omega)\!+\!\sigma)^2}\right\}.
\end{eqnarray}
is the kernel function and 
\begin{eqnarray}
\delta(\omega)=\mathcal{P}\int_0^{\infty}\dfrac{d\tilde\omega \rho(\tilde\omega)}{\omega\!-\!
\tilde\omega}\!
\label{Eq_Lamb_shift}
\end{eqnarray}
can be interpreted as the nonlinear Lamb shift of the cavity frequency $\omega_c$. Here $\mathcal{P}$ stands for the Cauchy principal value and $R_j$ is the contribution of poles (if at all existing)
\begin{eqnarray}
R_j=\dfrac{e^{(\sigma_j+i\omega_j)t}}{1-\Omega^2\int_0^{\infty}\dfrac{d\omega\rho(\omega)}{[\sigma_j+i(\omega_j+\omega)]^2  } }.
\end{eqnarray}
\section{Decay process from the steady state}
\label{App_Decay}
After applying a long rectangular pulse, both the cavity amplitude and spin ensemble settle to a finite value in the steady state (see Figs.~\ref{Figure_long_pulse_with_Lorentz},\ref{Fig_A_2_Jx_2_Lor},\ref{Fig_A_2_Jx_2_q_Gauss}). Here we explore the decay process from this steady state solution in more detail. To avoid cumbersome expressions we present, without loss of generality, the results for the resonant case only, $\omega_p=\omega_c=\omega_s$. To obtain a stationary solution, we set the time derivatives in Eqs.~(\ref{Eq_a_Volt}), (\ref{Eq_bk_Volt}) to zero, $\dot{A}(t)=\dot{B}_k(t)=0$, go to the continuous limit (in frequency) and finally derive the following expressions for the cavity amplitude and for the expectation values of the following collective spin operators, 
\begin{eqnarray}
\label{a_station}
\!\!\!\!\!\!\!\!\!\!\!\!\!A_{st}&\!=\!&\dfrac{\eta}{-\kappa+i\Omega^2
\bigintss_0^{\infty} \!\!  d\omega \dfrac{\rho(\omega)}{\omega-\omega_s-i \gamma}}, 
\\
\!\!\!\!\!\!\!\!\!\!\!\!\!J_x^{st}+iJ_y^{st}&\!=\!&\dfrac{\sum_k g_k B_k^{st}}{2\Omega}\!=\!\dfrac{i A_{st} \Omega}{2}\int\limits_0^{\infty} d\omega \dfrac{\rho(\omega)}{\omega-\omega_s-i \gamma}.
\end{eqnarray}
It can be easily proved, that the expressions above are real because the $q$-Gaussian is symmetric with respect to $\omega_s$, and as a consequence,  $J_y^{st}=0$ and $A_{st}\in {\mathbb R}$. Note that the second term in the Volterra equation (\ref{Eq_a_with_Bk0}) stands for the energy coming back to the cavity from the initial (steady) state of a spin ensemble, which in the continuous limit is found to be 
\begin{eqnarray}
&&\sum_k g_k B_k^{st} e^{-i(\omega_k-\omega_s-i\gamma)t}=
\\\nonumber
&& iA_{st}\Omega^2\int d\omega \rho(\omega) \dfrac{e^{-i(\omega_k-\omega_s-i\gamma)t}}{\omega-\omega_s-i\gamma},
\end{eqnarray}
leading to the following Volterra equation
\begin{eqnarray}
\nonumber
&&\!\!\!\!\!\!\!\!\!\dot A(t)\!=\!-\kappa A(t)\!+\!iA_{st}\Omega^2\int_0^{\infty} d\omega \rho(\omega) \dfrac{e^{-i(\omega-\omega_s-i\gamma)t}}{\omega-\omega_s-i\gamma}-
\\
&&\!\!\!\!\!\!\!\!\!\Omega^2 \int_0^{\infty} d\omega \rho(\omega) \int\limits_{0}^t d\tau
e^{-i(\omega-\omega_s-i \gamma)(t-\tau)}A(\tau).
\label{Eq_a_with_Bk0_cont_1}
\end{eqnarray}
From this expression we can conclude that the energy which is first stored and then released from the spin-ensemble is exactly the reason for the pronounced overshoot in the cavity amplitude (see the example shown in Fig.~\ref{Figure_long_pulse_with_Lorentz}a). Note that if initially the spin ensemble is in the ground state, $B_k(0)=0$, then the overshoot effect will never occur, as is the case for initial conditions described in Appendix \ref{App_Lapl_Transf} (the cavity is fed with a single photon and a spin ensemble is in the ground state).

Next, employing the Sokhotski-Plemelj theorem, in the limit of $\gamma\rightarrow 0$ 
\begin{eqnarray}
\nonumber
\int_0^{\infty}\dfrac{d\omega F(\omega)}{\omega-\omega_s-i \gamma}=
\mathcal{P}\int_0^{\infty}\dfrac{d\omega F(\omega)}{\omega-
\omega_s}+i\pi F(\omega_s),
\label{Sokhotski_theorem}
\end{eqnarray}
where $\mathcal{P}$ denotes the Cauchy principal value, Eqs.~(\ref{a_station}, \ref{Eq_a_with_Bk0_cont_1}) are finally reduced to (the resonance case, $\omega_p=\omega_c=\omega_s$, is considered only)
\begin{eqnarray}
\label{Ast_gen}
A_{st}=-\dfrac{\eta}{\kappa+\pi\Omega^2\rho(\omega_s)},
\end{eqnarray}
and
\begin{eqnarray}
\nonumber
&&\!\!\!\!\!\!\!\!\!\!\!\!\!\!\!\!\!\!\dot A(t)\!=\!-\kappa A(t)\!+
\\\nonumber
&&\!\!\!\!\!\!\!\!\!\!\!\!\!\!\!\!\!\!A_{st}\Omega^2
\left\{ \int_0^{\infty} d\omega \rho(\omega) \dfrac{\sin[(\omega-\omega_s)t]}{\omega-\omega_s}-\pi\rho(\omega_s)\right\}-
\\
&&\!\!\!\!\!\!\!\!\!\!\!\!\!\!\!\!\!\!\Omega^2 \int_0^{\infty} d\omega \rho(\omega) \int_{0}^t d\tau
e^{-i(\omega-\omega_s-i\gamma)(t-\tau)}A(\tau).
\label{Eq_a_with_Bk0_cont_2}
\end{eqnarray}
This equation describes the damped Rabi oscillations from the steady state after switching off a long pulse for a general form of the spin density, including both Lorentzian and $q$-Gaussian distributions [see results presented in Figs.~\ref{Figure_long_pulse_with_Lorentz}a),b)].

\end{document}